\documentclass[11pt,a4paper]{article}
\usepackage{jheppub}
\usepackage{pdflscape}
\usepackage{amsmath}
\usepackage{amssymb}
\usepackage{dcolumn}
\usepackage{bm}
\usepackage{color}
\usepackage{epsfig}
\usepackage{amsfonts}
\usepackage{graphicx}
\usepackage{subfigure}
\usepackage{dcolumn}

\newcommand{\be}{\begin{equation}}
\newcommand{\ee}{\end{equation}}
\newcommand{\bea}{\begin{eqnarray}}
\newcommand{\eea}{\end{eqnarray}}

\def\be{\begin{equation}}
\def\ee{\end{equation}}
\def\bea{\begin{eqnarray}}
\def\eea{\end{eqnarray}}

\begin{document}

\title{ Quark Stars in 4D Einstein-Gauss-Bonnet gravity with an Interacting Quark Equation of State}

\author[a]{Ayan Banerjee,}
\author[b]{Takol Tangphati,}
\author[c,d]{Daris Samart,}
\author[d,e,f,g]{and Phongpichit Channuie}

\affiliation[a]{Astrophysics and Cosmology Research Unit, University of KwaZulu Natal, Private Bag X54001, Durban 4000,
South Africa,}
\affiliation[b]{Department of Physics, Faculty of Science, Chulalongkorn University, \\Bangkok 10330, Thailand,}
\affiliation[c] {Department of Physics, Faculty of Science, Khon Kaen University, Khon Kaen, 40002, Thailand}
\affiliation[d] {School of Science, Walailak University, Nakhon Si Thammarat, 80160, Thailand} 
\affiliation[e] {College of Graduate Studies, Walailak University, Nakhon Si Thammarat, 80160, Thailand}
\affiliation[f] {Research Group in Applied, Computational and Theoretical Science (ACTS), \\Walailak University, Nakhon Si Thammarat, 80160, Thailand} 
\affiliation[g] {Thailand Center of Excellence in Physics, Ministry of Higher Education, Science, \\Research and Innovation, Bangkok 10400, Thailand}

\emailAdd{ayanbanerjeemath@gmail.com}
\emailAdd{takoltang@gmail.com}
\emailAdd{darisa@kku.ac.th}
\emailAdd{channuie@gmail.com}

\abstract{The detection of gravitational waves (GWs) from the binary neutron star (BNS) has opened a new window on the gravitational wave astronomy. With current sensitivities, detectable signals coming from compact objects like neutron stars turn out to be a crucial ingredient for probing their structure, composition, and evolution. Moreover, the astronomical observations on the pulsars and their mass-radius relations put important constraints on the dense matter equation of state (EoS). In this paper, we consider a homogeneous and unpaired charge-neutral $3$-flavor interacting quark matter with $\mathcal{O}(m_s^4)$ corrections that account for the moderately heavy strange quark instead of the naive MIT bag model. In this article, we perform a detailed analysis of strange quark star in the context of recently proposed $4D$ Einstein-Gauss-Bonnet (EGB) theory of gravity. However, this theory does not have standard four-dimensional field equations. Thus, we thoroughly show that the equivalence of the actions in the regularized $4D$ EGB theory and in the original one is satisfied for a spherically symmetric spacetime. We pay particular attention to the possible existence of massive neutron stars of mass compatible with $M \sim 2 M_{\odot}$. Our findings suggest that the fourth-order corrections parameter ($a_4$) of the QCD perturbation and coupling constant $\alpha$ of the GB term play an important role in the mass-radius relation as well as the stability of the quark star. Finally, we compare the results with the well-measured limits of the pulsars and their mass and radius extracted from the spectra of several X-ray compact sources.}

\keywords{4D EGB gravity; Interacting Quark Equation of State; Quark stars}

\maketitle

\section{Introduction}

Over the past few years there have been a lot of interest in higher
derivative gravity (HDG) theories. 
Although, many approaches have been introduced in order to modify GR and perhaps construct HDG theories appear in an effective level. In fact, this theory has been proposed in an expectation that higher order corrections to Einstein's GR might solve the singularity problem of black holes, avoids causality problems at the classical level and so on. Among the higher curvature gravity theories,  Lovelock gravity (LG) \cite{Lovelock,Lovelock:1972vz} has attracted considerable attention. Lovelock theory is the most general metric theory of gravity yielding conserved second order equations of motion in arbitrary number of dimensions $D$. In particular, Lovelock gravity is the natural generalization of Einstein gravity to higher dimensions and this theory coincides with Einstein theory in $D = 4$.
It follows that Lovelock gravities share a number of additional nice properties with Einstein gravity that are not enjoyed by other more general higher curvature theories. Most prominently, this theory is not only linear metric perturbations about flat spacetime but all perturbations are second-order, and not only about flat space, but any background. Consequently, Lovelock gravity  theories are free from many of the pathologies that plague general higher derivative gravity theories. 
Besides the Einstein-Hilbert term plus a cosmological constant, there is
a Gauss-Bonnet (GB) term $\mathcal{G}$ allowed in Lovelock's action in higher-dimensional spacetime.
This theory is called Einstein-Gauss-Bonnet (EGB) gravity \cite{Lanczos:1938sf},
 which appears in the low energy effective action of heterotic string theory \cite{Zwiebach:1985uq}. Other related works can be
found in Refs. \cite{Wiltshire:1985us,Boulware:1985wk,Wheeler:1986}.

However, in 4$D$, the Gauss-Bonnet (GB) term becomes
a topological invariant and  does not contribute to the gravitational dynamics. Recently, Glavan and Lin \cite{Glavan:2019inb} have proposed a dimensional regularization of the Gauss-Bonnet equations and obtain a 4$D$ metric theory that can bypasses the  conclusions of Lovelock's theorem and avoids Ostrogradsky instability. The approach has been formulated in $D$-dimensions, by rescaling the coupling constant $\alpha \to \alpha/(D -4)$, 
and then taking the limit $D \to 4$.  Thus, the GB term shows
a nontrivial contribution to the gravitational dynamics, which is referred to as the 4$D$ Einstein-Gauss-Bonnet (EGB) gravity. This process is  referred to as \textit{regularization}, which was first considered by Tomozawa \cite{Tomozawa:2011gp} 
with finite one-loop quantum corrections to Einstein gravity. As a result static spherically symmetric black hole solutions and their physical properties have been investigated, see \cite{Ghosh:2020syx,Konoplya:2020juj,Kumar:2020uyz,li04,Kumar:2020xvu,Zhang:2020sjh,Liu:2020vkh,we03,li05}. It was found that a rotating generalization has been studied in \cite{Kumar:2020owy,NaveenaKumara:2020rmi}  using the Newman-Janis algorithm. However, it is well known \cite{Hansen:2013owa} that the Newmann-Janis trick is not generally applicable in higher curvature theories. Thus, rotating solutions in 4$D$ EGB gravity remain to be found. Nevertheless, a number of interesting results in support of this ideas such as geodesics motion and shadow \cite{Zeng:2020dco,Guo:2020zmf}, the strong/ weak gravitational lensing by black hole  \cite{Islam:2020xmy,Kumar:2020sag,Heydari-Fard:2020sib,Jin:2020emq}, spinning test particle \cite{zh03}, thermodynamics AdS black hole \cite{sa03}, Hawking radiation \cite{Zhang:2020qam,Konoplya:2020cbv}, quasinormal modes \cite{Churilova:2020aca,Mishra:2020gce,ar04}, and wormhole solutions \cite{Jusufi:2020yus,liu20}, were extensively analyzed. Additionally, new quark stars in the context of 4$D$ Einstein-Gauss-Bonnet gravity have been recently proposed in Refs.\cite{Banerjee:2020stc,Banerjee:2020yhu} with various equations of state. Additionally, in the $4D$ EGB theory, one can facilitate the calculation of the Euclidean action of the vacuum bubble configurations that maybe be a technical advantage for this special theory \cite{Samart:2020sxj}. Many aspects of the 4$D$ EGB gravity were discussed in the literature, 
see \cite{Jusufi:2020qyw,Yang:2020jno,Ma:2020ufk,si20} for instance.

However, the validity of this 4$D$ EGB theory is at present under debate by several grounds.  Let us mention a few examples. It was demonstrated in Ref. \cite{Gurses:2020ofy} that there exists no four-dimensional equations of motion constructed from the metric alone that could serve as the equations of motion for such a theory. The four-dimensional theory must introduce additional degrees of freedom. In any case, it cannot be a pure metric theory of gravity. Moreover, from the perspective of scattering amplitudes, it was demonstrated that the limit leads to an additional scalar degree of freedom, confirming the previous analysis. Taking these issues together, the approach proposed by Glavan and Lin maybe incomplete. In other words, a description of the extra degree of freedom is required. Since then, several regularization schemes, e.g., see \cite{Lu:2020iav,Kobayashi:2020wqy,Fernandes:2020nbq}, have been proposed in order to overcome these shortcomings. In fact, Lu and Y. Pang \cite{Lu:2020iav} have shown that the Kaluza-Klein approach of the $D \to 4$ limit leads to a class of scalartensor theory that belongs to the Horndeski class. Thus, it is important to check the equivalence of the actions in the regularized and novel $4D$ EGB theory for the particular kind of static spherically symmetric spacetime.

Neutron stars (NSs) are dense, compact astrophysical objects which  are the remains of very massive stars  (10–30 $M_{\odot}$) that ended their lives in supernova explosions \cite{Woosley,Heger:2002by}.  However the discovery of neutron stars with masses around 
2 $M_{\odot}$ \cite{Demorest:2010bx,Antoniadis:2013pzd}, put forward a strong on the EoS of matter
in neutron stars. But in the interior of these objects determining the true state of the matter  is still an open question, which is the greatest importance for
particle physics as well as stellar astrophysics alike. Moreover, the composition and  structure of compact stars depend on the nature of strong interaction. Under such conditions, the presence of  different exotic matter with large strangeness fraction such as hyperon matter, Bose-Einstein condensates of strange mesons and quark matter may occur in neutron star interior. Other theories suggest that each exotic component of dense matter makes the EoS soft, and  soft EoS generally gave rise to a compact star with smaller maximum mass and radius than
those of a stiffer EoS \cite{Lattimer:2000nx}.

However, the mass measurements of the  massive neutron star J0348+0432 \cite{Antoniadis:2013pzd} with $2.01 \pm 0.04 M_{\odot}$, and PSR J1614-2230 \cite{Zhang:2019fog} with $M=1.97 \pm 0.04 $M$_\odot$, has set rigid constraints on the theoretical models of dense nuclear matter. The existence of such massive stars has important implications for dense matter in  Quantum Chromodynamics (QCD), where a phase transition from hadronic matter to a deconfined quark phase should occur in neutron star interior.  Even more intriguing the existence of a quark core in a neutron star, is the possible existence of a new family of compact stars composed of the three lightest quark flavor states (up, down, and  strange quarks) satisfying the Bodmer-Witten hypothesis \cite{Witten,Bodmer}. Despite all the advances in our understanding of QCD, most of the analysis
for quark stars still continues to be performed in the context of the MIT bag model \cite{Chodos:1974pn,Peshier,Chodos:1974je}. In MIT bag model, quarks in the bag are considered as a free Fermi gas and provide mechanism of quark confinement. 

However, MIT bag model has some limitations that violates chiral symmetry even in the limit of massless quark. Moreover it was found that this EOS is not sufficiently reliable to characterize a system with interacting quarks or more complex structures.  Thus, some authors have suggested some modified models, for example, the three-flavor 
quark matter with the particular symmetry is called the color-flavor locked (CFL)  matter \cite{Alford:1998mk}. It is widely believed that the CFL matter is a real ground state of QCD at asymptoticly large densities \cite{Alford:1997zt}. However, at extremely high density, the phase of matter is less certain. In these proceedings a 2-component model for quark stars have been reported \cite{Asbell:2017zxp} that can produce stars as heavy as $2M_{\odot}$. 

Motivated by the newly proposed EoS which is homogeneously confined in the stellar interior with 3-flavour neutral charge and a fixed strange quark mass \cite{Asbell:2017zxp}, we propose a simple model for  quark star in 4$D$ EGB gravity. This accumulation will lead to various changes in  the mass-radius relation of a quark star whose results were compared with compact stars candidates like J0348+0432 \cite{Antoniadis:2013pzd}, PSR J1614-2230 \cite{Zhang:2019fog}, J1903+0327 \cite{Freire}, 4U 1608-52 \cite{Guver:2008gc}. This paper is arranged as follows: In Sec. \ref{sec20}, we take a short recap of the  regularized 4$D$ EGB gravity, as presented
in \cite{Fernandes:2020nbq}. Then the equivalence of the actions in the regularized and novel $4D$ EGB theory
 has been proved for a static spherically symmetric spacetime.  Sec. \ref{sec2} is devoted to summarize
the original formulation of $4D$ EGB theory and the associated gravitational field equations in static, 
spherically symmetric context.  In the same section, we show that the trace of fields equation derived for the original 4D theory is exactly the same as of the regularized one. In continuation we also derive the  field equations describing the structure of the relativistic stars in $4D$ EGB gravity.
In Sec. \ref{sec3} we introduce a QCD motivated EoS. In Sec. \ref{sec4}, we perform a detailed numerical analysis and present mass-radius relations for quark matter stars by solving the customized TOV equations. We demonstrate the physical properties of a constructed quark star in Sec. \ref{sec5}. Finally, we summarize our findings and discuss our results in Sec. \ref{sec6}.

\section{Regularized scheme of 4D Gauss-Bonnet gravity and the field equation : a short recap}
\label{sec20}
In the present section, we take a short recap of the regularization scheme in order to find the regularized action which is free from divergences, and produces well behaved second-order
field equations that can be regularly used for gravitational
physics. Let us start from the general action of EGB theory in $D$-dimensional space-time and also derive the equations of motion. The action takes the form
\begin{eqnarray}
 S= \frac{c^4}{16\pi G_D} \int_{\mathcal{M}}d^D x \sqrt{-g} \left(R + \hat\alpha \mathcal{L}_{\text{GB}} \right) + \mathcal{S}_{\text{matter}},
\end{eqnarray}
where $g$ denotes the determinant of the metric $g_{\mu\nu}$ and $\alpha$ is the Gauss-Bonnet coupling constant. As also noted that
 $\mathcal{S}_{\text{matter}}$ is the action associated with matter field, and the Gauss-Bonnet term is
\begin{eqnarray}\label{gravity field eq}
\mathcal{L}_{\text{GB}}= R^2-4R_{\mu\nu}R^{\mu\nu}+R_{\mu\nu\alpha\beta}R^{\mu\nu\alpha\beta}~.
\end{eqnarray}
Using the standard technique, we take a variation of the above action with respect to the metric $g_{\mu\nu}$ to obtain the field equation $\hat{\alpha} \to \frac{\alpha}{(D-4)}$ in the $D\to 4$ limit:
\begin{equation}\label{GBeq}
G_{\mu\nu} + \alpha H_{\mu\nu} = \frac{8 \pi G}{c^{4}} T_{\mu\nu}\,\,{\rm where}\,\,T_{\mu\nu}= -\frac{2}{\sqrt{-g}}\frac{\delta\left(\sqrt{-g}\mathcal{S}_m\right)}{\delta g^{\mu\nu}},
\end{equation}
with $G_{\mu\nu}$ is the Einstein tensor and $H_{\mu\nu}$ is a tensor carrying the contributions from
the Gauss-Bonnet (GB) term given by
\begin{eqnarray}
G_{\mu\nu} &=& R_{\mu\nu}-\frac{1}{2}R~ g_{\mu\nu},\nonumber\\
H_{\mu\nu} &=& 2\Bigr( R R_{\mu\nu}-2R_{\mu\sigma} {R}{^\sigma}_{\nu} -2 R_{\mu\sigma\nu\rho}{R}^{\sigma\rho} + R_{\mu\sigma\rho\delta}{R}^{\sigma\rho\delta}{_\nu}\Bigl)- \frac{1}{2}~g_{\mu\nu}~\mathcal{L}_{\text{GB}},\label{FieldEq}
\end{eqnarray}
where $R_{\mu\nu}$ is the Ricci tensor, $R$ and $R_{\mu\sigma\nu\rho}$ are the  Ricci scalar and the Riemann tensor, respectively. A first and simple consistency check that can be done is to verify whether or not the obtained solutions satisfy the trace of the Gauss-Bonnet field equations when the $D\rightarrow 4$ limit is taken. As a specific example of the static spherically symmetric metic (\ref{metric01}), we consider a  compact stellar object. Here, for the sake of simplicity, we assume that the energy momentum tensor $T_{\mu\nu}$ is a perfect fluid matter source and describes the interior of a star, which in
this study is written as  
\begin{eqnarray}
T_{\mu\nu} = (\epsilon+P)u_{\nu} u_{\nu} + P g_{\nu \nu}, \label{em}
\end{eqnarray}
where $P=P(r)$ is the pressure, $\epsilon \equiv \epsilon(r)$ is the energy density of matter, and $u_{\nu}$ is a $D$-velocity. In this paper, we will start from a static spherically symmetric $D$-dimensional metric anstaz with two independent functions of the radial coordinate which takes the form:   
\begin{eqnarray}\label{metric01}
    ds^2_{D} = - W(r) dt^2 + H(r) dr^2 + r^{2}d\Omega_{D-2}^2, 
\end{eqnarray} 
where $d\Omega_{D-2}^2$ is the metric on the unit $(D-2)$-dimensional sphere and $W=W(r)$ and $H=H(r)$ are functions of $r$, solely. Invoking the metric (\ref{metric01}) with the energy momentum tensor (\ref{em}), in the limit $D \to 4$,  the $(t,t)$ and $(r,r)$ components of the field equation yield
\begin{eqnarray}
   \frac{8 \pi G}{c^{4}} \epsilon &=& \frac{1}{r^2} + \frac{1}{r H} \left( \frac{H'}{H} - \frac{1}{r} \right) 
   - \frac{\alpha (H - 1)}{r^4 H^3}\,\Big(H^2 -H - 2H' r\Big)\,, \label{DR1} \\ 
  \frac{8 \pi G}{c^{4}}  P &=& -\frac{1}{r^2} + \frac{1}{r H} \left( \frac{1}{r} + \frac{W'}{W} \right) 
  +\frac{\alpha (H - 1)}{r^4 H^2 W}\,\Big(W(H - 1) + 2 W' r\Big) \,,\label{DR2}
\end{eqnarray}
where primes denote derivative with respect to $r$. The basic motivation of this theory is to construct a nontrivial theory by redefining the GB coupling constant as $\hat{\alpha} \to \frac{\alpha}{(D-4)}$
\cite{Glavan:2019inb}. This, however, turns out to be questionable in numerous aspects. Even though there are 
several criticisms against this model, including the above limiting procedure being invalid (see references \cite{Ai:2020peo,Gurses:2020ofy,Lu:2020iav,Kobayashi:2020wqy, Hennigar:2020lsl,Fernandes:2020nbq,shu}). 
To address this issue, some regularized $4D$ EGB theories were proposed, such as the Kaluza–Klein-reduction 
procedure \cite{Lu:2020iav,Kobayashi:2020wqy}, the conformal subtraction procedure \cite{Hennigar:2020lsl,Fernandes:2020nbq}, 
and ADM decomposition analysis \cite{Aoki:2020iwm}. This leads to the fact that regularization
procedure is not unique.  Here, we follow the approach as proposed in \cite{Fernandes:2020nbq,Lu:2020iav,Yang:2020jno} to yield a resulting divergent free action
\begin{eqnarray}\label{action1}
 S &=&  \frac{c^4}{16\pi G}\int_{\mathcal{M}} \mathrm{d}^4 x \sqrt{-g} \Big[R+ \alpha \Big(4G^{\mu \nu}\nabla_\mu \phi \nabla_\nu \phi - \phi \mathcal{L}_{\text{GB}}+ 4\Box \phi (\nabla \phi)^2 + 2(\nabla \phi)^4 \Big] + \mathcal{S}_{\text{matter}},
\end{eqnarray}
where $\phi$ is an extra scalar gravitational degree-of-freedom  inherent from $D$ dimensions. While it is true that the GB term is topological in $D=4$, what this means is that its variation vanishes. In other words, a pure GB contribution to the action yields a trivial contribution to the field equations. This is not the situation in (\ref{GB3}), though. Effectively, the scalar has acted as a Lagrange multiplier in the action allowing for the GB term itself to appear in the four-dimensional field equations. Note that $L_{GB}$ does not vanish in $D=4$, this means it does in fact make a contribution to the field equations. Interestingly, this action belongs to a subclass of the Horndeski gravity \cite{Horndeski:1974wa,Kobayashi:2019hrl}  with  $G_2=8 \alpha X^2-2\Lambda_0$, $G_3=8 \alpha X$, $G_4=1+4 \alpha X$ and $G_5 = 4 \alpha \ln X$ \big(where $X=-\frac{1}{2} \nabla_{\mu} \phi \nabla^{\mu} \phi$\big).
It was introduced by Kaluza–Klein reduction of the metric \cite{Lu:2020iav,Kobayashi:2020wqy}
\begin{eqnarray}
ds_D^2=ds_4^2+e^{2\phi}d\Omega_{D-4}^2,
\end{eqnarray}
or by conformal subtraction \cite{Hennigar:2020lsl,Fernandes:2020nbq}, where the subtraction background is defined under a conformal transformation ${\tilde g}_{ab} = e^{2\phi}g_{ab}$ and a counterterm, i.e., $-\alpha\int_{\mathcal{M}} \mathrm{d}^4\sqrt{-{\tilde g}}\,\tilde{\mathcal{G}}$, is added to the original action \cite{Fernandes:2020nbq}. Here, the scalar $\phi$ depends only on the external 4-dimensional coordinates, $ds^{2}_{4}$ is the 4-dimensional line element, and $d\Omega_{D-4}^2$ is the line element of the internal
maximally symmetric space. Varying the action (\ref{action1}), we obtain the equations of motion \cite{Fernandes:2020nbq}
\begin{eqnarray}\label{gravity field eq}
 G_{\mu \nu} =  \alpha \hat{\mathcal{H}}_{\mu \nu} + \frac{8\pi G}{c^4}T_{\mu \nu},
 \label{regularized-EFE}
\end{eqnarray}
where $T_{\mu \nu}$ is the energy-momentum tensor of the matter field as defined in (\ref{em}). In these geometries, the $\hat{\mathcal{H}}_{\mu\nu}$ is given by 
\begin{eqnarray}\label{H}
\hat{\mathcal{H}}_{\mu\nu} &=& 2R\big(\nabla_\mu \nabla_\nu \phi - \nabla_\mu\phi \nabla_\nu \phi\big) + 2G_{\mu \nu}\Big(\big(\nabla \phi\big)^2-2\Box \phi\Big)+ 4G_{\nu \alpha} \big(\nabla^\alpha \nabla_\mu \phi -\nabla^\alpha \phi \nabla_\mu \phi\big)\nonumber\\
 &+& 4G_{\mu \alpha} \big(\nabla^\alpha \nabla_\nu \phi - \nabla^\alpha \phi \nabla_\nu \phi\big) + 4R_{\mu \alpha \nu \beta}\big(\nabla^\beta \nabla^\alpha \phi - \nabla^\alpha \phi \nabla^\beta\phi\big) + 4\nabla_\alpha\nabla_\nu \phi \big(\nabla^\alpha \phi \nabla_\mu \phi\nonumber\\
 &-& \nabla^\alpha \nabla_\mu \phi \big)+4 \nabla_\alpha \nabla_\mu \phi \nabla^\alpha\phi \nabla_\nu \phi - 4\nabla_\mu \phi \nabla_\nu \phi \Big(\big(\nabla \phi\big)^2+ \Box \phi \Big) \nonumber\\&+&4\Box \phi\nabla_\nu \nabla_\mu \phi - g_{\mu \nu} \Big[ 2R\big(\Box \phi  - (\nabla \phi)^2\big)
\nonumber\\
&+& 4 G^{\alpha \beta} \big( \nabla_\beta \nabla_\alpha \phi -\nabla_\alpha \phi \nabla_\beta \phi \big)+ 2\big(\Box \phi \big)^2 
- \big( \nabla \phi\big)^4 \nonumber\\&+& 2\nabla_\beta \nabla_\alpha\phi \big(2\nabla^\alpha \phi \nabla^\beta \phi - \nabla^\beta \nabla^\alpha \phi \big) \Big].
\end{eqnarray}
Hence, by varying with respect to the scalar field, we get
\begin{eqnarray}\label{scalar field}
\frac{1}{8}\mathcal{L}_{\text{GB}} &=&  R^{\mu \nu} \nabla_{\mu} \phi \nabla_{\nu} \phi - G^{\mu \nu}\nabla_\mu \nabla_\nu \phi - \Box \phi (\nabla \phi)^2 +(\nabla_\mu \nabla_\nu \phi)^2- (\Box \phi )^2 \nonumber\\&-& 2\nabla_\mu \phi \nabla_\nu \phi \nabla^\mu \nabla^\nu \phi\,.
\end{eqnarray}

Following \cite{Fernandes:2020nbq}, one can obtain the 
trace of the field equations (\ref{gravity field eq}) to satisfy
\begin{eqnarray}\label{GB3}
R+\frac{\alpha}{2} \mathcal{L}_{\text{GB}}= -\frac{8\pi G}{c^4}T.
\end{eqnarray}

In \cite{Fernandes:2020nbq}  it is claimed that the trace of the field
equation is exactly same form as of the original $4D$ EGB theory. The present approach exactly
reproduces the well-defined field equation for
the $4D$ EGB theory, and implies a hidden scalar degree of freedom in the original
theory. Note that the scalar field equation can be seen
to be exactly equivalent to the condition $\tilde{\cal G}=0$. This means that the counter term added to the action must vanish 
on-shell. In other words, the on-shell action of the regularized theory takes exactly the same form of the original theory, and
that the classical evolution of the gravity-matter system
is independent of the hidden scalar field \cite{Fernandes:2020nbq}. In the spherical coordinates, it is rather straightforward to calculate non-vanishing components of the gravitational field equations. With the on-shell condition, the scalar field equation is an identity and then can be eliminated.

By taking into account the metric (\ref{metric01}), a straightforward calculation in the limit $D\rightarrow4$ gives the nonvanishing components of the gravitational field equations written in terms of $W(r)$ and $H(r)$. As we have mentioned before, if the on-shell action of the regularized $4D$ EGB theory equals to the one of the
novel $4D$ EGB theory in a static and spherically symmetric
spacetime (\ref{metric01}), both of them will give the same solutions. From Eq. (\ref{scalar field}), we can verify that the on-shell condition does not involve the matter content. 
Having used the line element (\ref{metric01}), the $(t,t)$ and $(r,r)$ components of the field equation (\ref{regularized-EFE}) are:
\begin{eqnarray}
   \frac{8 \pi G}{c^{4}} \epsilon &=& \frac{1}{r^2} + \frac{1}{r H} \left( \frac{H'}{H} - \frac{1}{r} \right) \nonumber\\
   &+& \alpha  \Bigg(\frac{6 H'\phi'}{r^2 H^3}-\frac{2 H' \phi'}{r^2 H ^2}+\frac{H' W' \phi'^2}{H^3 W }+\frac{2 H'\phi'^3}{H^3}+\frac{6 H'\phi'^2}{r H ^3}+\frac{4\phi''}{r^2 H}-\frac{4 \phi''}{r^2 H^2}\nonumber\\
   &&\qquad -\,\frac{2 \phi'^2}{r^2 H } -\frac{2 \phi'^2}{r^2 H^2}-\frac{4 W'\phi'^2}{r H^2 W }-\frac{W'^2 \phi'^2}{H ^2 W^2} -\frac{2 W'\phi'\phi''}{H^2 W }+\frac{\phi'^4}{H^2}-\frac{4 \phi'^2 \phi''}{H^2}-\frac{8 \phi'\phi'' }{r H^2}\Bigg)\,, \label{tt-phi} \\ 
  \frac{8 \pi G}{c^{4}}  P &=& -\frac{1}{r^2} + \frac{1}{r H} \left( \frac{1}{r} + \frac{W'}{W} \right) \nonumber\\
  &+&\alpha  \Bigg(\frac{2 H' \phi '^3}{H^3}-\frac{H' W' \phi'^2}{H^3 W}+\frac{H'^2 \phi'^2}{H^4}-\frac{4 H' \phi '^2}{r H^3}-\frac{4 H' \phi' \phi''}{H^3}+\frac{6 W' \phi'}{r^2 H^2 W}-\frac{2 W' \phi '}{r^2 H W}\nonumber\\
  &&\qquad -\,\frac{2 \phi '^2}{r^2 H}+\frac{6 \phi'^2}{r^2 H^2}+\frac{6 W' \phi'^2}{r H^2 W}+\frac{2 W' \phi' \phi''}{H^2 W}+\frac{4 \phi''^2}{H^2}+\frac{3 \phi '^4}{H^2}-\frac{4 \phi'^2 \phi''}{H^2}+\frac{8 \phi ' \phi ''}{r H^2}\Bigg) \,.\label{rr-phi}
\end{eqnarray}
In order to demonstrate an equivalent between the regularized 4D EGB gravity and the novel $4D$ EGB theory, one must quantify the solution of $\phi$ making both of theories to have the same solution. We demonstrate this equivalence by subtracting Eqs. (\ref{DR1}) and (\ref{tt-phi}) for the $(tt)$ component and Eqs. (\ref{DR2}) and (\ref{rr-phi}) for the $(rr)$ part. Thus, we find
\begin{eqnarray}
0 &=& \frac{2 H'}{r^3 H^3}-\frac{2 H'}{r^3 H^2}+\varphi  \left(\frac{6 H'}{r^2 H^3}-\frac{2 H'}{r^2 H^2}\right)\nonumber\\&+&\varphi ^2 \left(\frac{H' W'}{H^3 W}+\frac{6 H'}{r H^3}-\frac{2}{r^2 H}-\frac{2}{r^2 H^2}-\frac{W'^2}{H^2 W^2}-\frac{4 W'}{r H^2 W}\right)\nonumber\\
&+& \frac{2 \varphi ^3 H'}{H^3}-\frac{2}{r^4 H}+\frac{1}{r^4 H^2}+\varphi ' \left(\frac{4}{r^2 H}-\frac{4}{r^2 H^2}-\frac{4 \varphi ^2}{H^2} -\varphi  \left[\frac{2 W'}{H^2 W}+\frac{8}{r H^2}\right]\right)\nonumber\\&+&\frac{\varphi ^4}{H^2}+\frac{1}{r^4}\,,\label{varphi-tt}
\\
0 &=& \varphi ^2 \left(\frac{H'^2}{H^4} -\frac{H' W'}{H^3 W}-\frac{4 H'}{r H^3}-\frac{2}{r^2 H}+\frac{6}{r^2 H^2}+\frac{6 W'}{r H^2 W}\right)+\frac{2 \varphi ^3 H'}{H^3}\nonumber\\&+&\varphi ' \left(\varphi  \left[\frac{2 W'}{H^2 W}-\frac{4 H'}{H^3}+\frac{8}{r H^2}\right]-\frac{4 \varphi ^2}{H^2}\right)\nonumber\\
&+&\frac{2}{r^4 H}-\frac{1}{r^4 H^2}+\frac{2 W'}{r^3 H^2 W}-\frac{2 W'}{r^3 H W}+\varphi  \left(\frac{6 W'}{r^2 H^2 W}-\frac{2 W'}{r^2 H W}\right)\nonumber\\&+&\frac{4 \varphi '^2}{H^2}+\frac{3 \varphi ^4}{H^2}-\frac{1}{r^4}\,,
\label{varphi-rr}
\end{eqnarray}
where the new variable, $\varphi$, is defined as $\varphi = \phi'$ and $\varphi' = \phi''$. We observe that the matter parts of $\epsilon(r)$ and $P(r)$ do not involve with the solution of $\varphi$. Furthermore, one may rewrite the Eqs. (\ref{varphi-tt}) and (\ref{varphi-rr}) by eliminating $\varphi'$, and this gives
\begin{eqnarray}
0 &=& \left(\sqrt{H}- r \varphi - 1\right) \Bigg(r^3 H W'^2\varphi -r W' W \left[H' r^2 \varphi + 2 H^{3/2} r \varphi +2 H^2-2 H (3 r \varphi +1)\right]]\nonumber\\
&&\quad\quad\quad\quad\quad\quad\quad\quad\; +  \left(\sqrt{H}+ r \varphi + 1\right) W^2\big[ H^2 - 8 \sqrt{H}^{3} +H \left(2 r \varphi-r^2 \varphi ^2 +7\right)\nonumber\\
&&\quad\quad\quad\quad\quad\quad\quad\quad\;+2 H' \sqrt{H} r-2 H' r (r \varphi +2) \big] \Bigg)\,,
\label{sol-varphi-tt}\\
0 &=& \frac{H-(r \varphi +1)^2}{\Big(W' r^2 \varphi +W \big[2 (r \varphi +1)^2-2 H\,\big]\Big)^2}\nonumber\\
&\times& \Bigg(W^3 \Big[H-(r \varphi +1)^2\Big] \Big[H^2 \left(4 H' r-9 r^4 \varphi ^4-12 r^3 \varphi ^3-10 r^2 \varphi ^2-4 r \varphi +3\right)
-4 H'^2 r^2 \nonumber\\
&&\quad\quad\; +\,4 H' H r \left(r^2 \varphi ^2-2 r \varphi -1\right)+3 H^4+H^3 \left(6 r^2 \varphi ^2+4 r \varphi -6\right)\Big]\nonumber\\
&&\quad\quad\; +\,2 r^3 H W'^2 W \varphi \Big[ + \left(3 r^3 \varphi ^3+6 r^2 \varphi ^2+11 r \varphi +4\right)-H' r^2 \varphi -H^2 (3 r \varphi +4) \Big]\nonumber\\
&&\quad\quad\; +\,2 r H W' W^2 \Big[H \Big(6 r^5 \varphi ^5+20 r^4 \varphi ^4+33 r^3 \varphi ^3 -2 r^2 \varphi  (H'-19 \varphi ) +19 r \varphi +4\Big)]\nonumber\\
&&\quad\quad\; +2 H' r^2 \varphi +H^3 (3 r \varphi +4-\,H^2 \left(9 r^3 \varphi ^3+22 r^2 \varphi ^2+22 r \varphi +8\right)\Big] ]\nonumber\\
&&\quad\quad\; +2r^5 H^2  W'^3 \varphi ^2\Bigg)\,.
\label{sol-varphi-rr}
\end{eqnarray}
Remarkably, one could easily verify that the above field equations (\ref{sol-varphi-tt}) and (\ref{sol-varphi-rr})
are identities when $\big(\sqrt{H}- r \varphi - 1\big)$ equals to zero. From the above equations, we know that it requires the scalar field to satisfy
\begin{eqnarray}
\varphi = \frac{1}{r}\Big( \sqrt{H} -1 \Big)\,.\label{varphi-sol}
\end{eqnarray}
Noting that the solutions of the spherically symmetric spacetime, the regularized $4D$ EGB theory will give the same solutions as of novel $4D$ EGB theory by substituting the solution of $\varphi(r)$ in (\ref{varphi-sol}) to the field equation in (\ref{regularized-EFE}). Finally, we can conclude that the two $4D$ EGB theories are equivalent in the static spherically symmetric metric (\ref{metric01}).Note that Eq.(\ref{varphi-sol}) proves that the two regularizations are equivalent. However, it seems plausible that Eq.(\ref{varphi-sol}) is not the only consistent scalar configuration. In other words, solutions other than Eq.(\ref{varphi-sol}) could exist that represents a completely novel quark star solutions. We will leave this interesting point for future study

Interestingly, the solution of $\varphi(r)$ in (\ref{varphi-sol}) found in this work takes the form like the typical scalar gravitational potential, i.e., $\varphi(r) = \varphi_0(r)/r$ with $\varphi_0(r) \equiv \sqrt{H(r)} - 1$ which is reasonable solution of the scalar field in the spherical symmetric spacetime. However, the scalar field should be regular at the origin $r=0$. To regulate the field $\varphi(r)$ at $r=0$, we therefore apply the Taylor-series expansion
 about the point $r=0$ for small $r$,  
\begin{eqnarray}
H(r) \simeq h_0 + h_1\,r + h_2\,r^2 + \mathcal{O}\left( r^3\right)\,,
\end{eqnarray}
where $h_{0,1,2,\cdots}$ are arbitrary constants. Without loss of generality, we can assume $h_0$ =1. Having used the binomial approximation to $\sqrt{H(r)}$\,, one finds for the small $r$ region, 
\begin{eqnarray}
\varphi(r) \simeq \frac12 h_1 \,.
\end{eqnarray}
We can clearly see that the $\varphi(r)$ solution is regular at $r \to 0$ limit, and therefore it does not divert near the core of the star.

\section{TOV equations in $4D$ EGB gravity}
\label{sec2}
We have revisited the regularized 4$D$ EGB framework and its reduction to the $4D$ EGB theory in the previous section. One could check that the resultant field equations given below Eqs. (\ref{DRE1}-\ref{DRE3}) of the novel 4$D$ EGB theory following a procedure present in Ref. \cite{Lin:2020kqe}. Therefore, the two 4$D$ EGB theories are equivalent in the static spherically symmetric spacetime. In other words, the solutions of the static spherically symmetric metric of the novel 4D EGB theory are also
solutions of the regularized one. 

Therefore, we anticipate using novel $4D$ EGB gravity not to constitute an impasse for our model in the present work. Here we start by considering the general action of Einstein-Gauss-Bonnet (EGB) gravity in $D$-dimensions and also deriving the equations of motion of the underlying theory. The action takes the form
\begin{equation}\label{action}
	\mathcal{S}_{\rm EGB}=\frac{c^4}{16 \pi G_{D}}\int d^{D}x\sqrt{-g}\Big[ R +\alpha\mathcal{L}_{\rm GB} \Big]
+\mathcal{S}_{\rm m},
\end{equation}
where $g$ denotes the determinant of the metric $g_{\mu\nu}$ and $\alpha$ is the Gauss-Bonnet coupling constant, $G_{D}$ is the $D$-dimensional Newton’s gravitational constant and $\mathcal{S}_{\rm m}$ is the matter field action. The Einstein-Gauss-Bonnet Lagrangian $\mathcal{L}_{\text{GB}}$ is given by
\begin{equation}
\mathcal{L}_{\text{GB}}=R^{\mu\nu\rho\sigma} R_{\mu\nu\rho\sigma}- 4 R^{\mu\nu}R_{\mu\nu}+ R^2\label{GB}.
\end{equation}
Note that adding the matter action $\mathcal{S}_{\text{m}}$ induces the energy momentum tensor $T_{\mu\nu}$ as shown in the previous section. A straightforward calculation is to consider the equation of motion on Eq.(\ref{GBeq}). We find
\begin{eqnarray}
g^{\mu\nu}\,\Big(G_{\mu\nu}+\alpha H_{\mu\nu}\Big) &=& -R + \alpha\big(2D-8\big) R_{\mu\nu}R^{\mu\nu}+\alpha\big(2-D/2\big)R^{2}+\alpha \big(2-D/2\big)R_{\mu\nu\sigma\rho}R^{\mu\nu\sigma\rho}\nonumber\\&=& -R-\frac{\alpha (D-4)}{2}\mathcal{L}_{\rm GB}.\label{FieldEqT}
\end{eqnarray}
Hence, one can clearly see that the GB term has no effect on
gravitational dynamics in four dimensions. However, if we
redefining the GB coupling
constant as $\alpha\rightarrow\alpha/(D-4)$ the trace of the field equation (\ref{GBeq}) yields
\begin{eqnarray}\label{GB3}
R+\frac{\alpha}{2}\mathcal{L}_{\rm GB} = -\frac{8 \pi G}{c^{4}}T\,,
\end{eqnarray}
which is exactly the same form as the trace of the field
equations of the regularized $4D$ EGB theory, as presented
in Eqs. (\ref{gravity field eq}) and (\ref{GB3}). Thus, the multiplicative factor of $(D -4)$ would be precisely cancelled by the proposed rescaling of the coupling constant $\alpha$, which would leave a nonvanishing contribution to the trace of the field equations as $D \to 4$. In view of the above circumstance it comes clear that two theories are equivalent in the static and spherically symmetric spacetime. In other words, the solutions of the spherically symmetric spacetime,  the regularized $4D$ EGB theory will give the same solutions as of the novel $4D$ EGB theory. In recent Refs. \cite{Lin:2020kqe}, authors have demonstrated that solutions of the static cylindrically symmetric
metric of the novel $4D$ EGB theory are also solutions of the regularized one. For solution describing stellar objects,
 we use the regularization process (see Refs. \cite{Glavan:2019inb,Cognola:2013fva}) in which the spherically symmetric solutions are also exactly same as those of other regularised theories \cite{Lu:2020iav,Hennigar:2020lsl,Casalino:2020kbt,Ma:2020ufk}.
 
In order to derive the TOV equation, it is convenient to introduce the line element of the compact object (neutron star) in the static spherical symmetry in the following form,
\begin{eqnarray}
ds^2 &=& - e^{2\Phi(r)}c^{2}dt^2 + e^{2\Lambda(r)}dr^2 + r^{2}d\Omega^2\,.
\label{line-exp}
\end{eqnarray}
Noting that the above line element is equivalent to the line element in (\ref{metric01}) for $e^{2\Phi(r)}c^{2} = W(r)$ and $e^{2\Lambda(r)}=H(r)$. We recall the $(tt)$, $(rr)$ and hydrostatic continuity equations (\ref {GBeq}) yield
\begin{eqnarray}\label{DRE1}
&& \frac{2}{r} \frac{d\Lambda}{dr} = e^{2\Lambda} ~ \left[\frac{8\pi G}{c^4} \epsilon - \frac{1-e^{-2\Lambda}}{r^2}\left(1-  \frac{\alpha(1-e^{-2\Lambda})}{r^2}\right)\right]\left[1 +  \frac{2\alpha(1-e^{-2\Lambda})}{r^2}\right]^{-1}, \\ 
&& \frac{2}{r} \frac{d\Phi}{dr} = e^{2\Lambda} ~\left[\frac{8\pi G}{c^4} P + \frac{1-e^{-2\Lambda}}{r^2} \left(1- \frac{\alpha(1-e^{-2\Lambda})}{r^2} \right) \right] \left[1 +  \frac{2\alpha(1-e^{-2\Lambda})}{r^2}\right]^{-1},\label{DRE2} \\
&& \frac{dP}{dr} = - (\epsilon + P) \frac{d\Phi}{dr}.  \label{DRE3}
\end{eqnarray} 
As usual, the asymptotic flatness implies $\Phi(\infty)=\Lambda(\infty)=0$ and the regularity at the center imposes the condition $\Lambda(0)=~0$.
Here we can define the gravitational mass within the sphere of radius $r$ given by $e^{-2\Lambda}= 1-\frac{2G m(r)}{c^2 r}$. It is straightforward to derive the Tolman-Oppenheimer-Volkoff (TOV) equations based on the underlying $4D$ EBG theory and we write by using (\ref{DRE1}-\ref{DRE3})
\begin{equation}
{dP \over dr} = -{G\epsilon(r) m(r) \over c^{2}r^2}\frac{\left[1+{P(r) \over \epsilon(r)}\right]\left[1+{4\pi r^3 P(r) \over c^{2}m(r)}-{2G\alpha m(r) \over c^{2}r^3}\right]}{\left[1+{4G\alpha m(r) \over c^{2}r^3}\right]\left[1-{2Gm(r) \over c^{2}r}\right]}. \label{e2.11}
\end{equation}
Notice that if we take the $\alpha \to 0$ limit, the above equation reduces to the TOV equation of the standard GR. From the last equality of Eq. (\ref{DRE1}), we obtain the gravitational mass 
\begin{equation}
m'(r)=\frac{6  \alpha G m(r)^2+4 \pi  r^6 \epsilon (r)}{4 \alpha G r m(r)+c^2 r^4}, \label{e2.12}
\end{equation}
We will use the initial condition of $m(r)$ such that $m(r=0)=0$. It is more convenient to work with the dimensionless variables. Therefore in the present analysis, we take $P(r)=\epsilon_{0}{\bar P}(r)$ and $\epsilon(r)=\epsilon_{0}{\bar \epsilon}(r)$ and $m(r)=M_{\odot}{\bar M}(r)$, with $\epsilon_{0}=1\,{\rm MeV}/{\rm fm}^{3}$. As a result, the above two equations become 
\begin{eqnarray}
{d{\bar P}(r) \over dr} = - \frac{c_1 {\bar \epsilon}(r) {\bar M}(r)}{r^2} \frac{\left[1+{{\bar P}(r) \over {\bar \epsilon}(r)}\right] \left[1+{c_2 r^3 {\bar P}(r) \over {\bar M}(r)}-{2 c_1 \alpha {\bar M}(r) \over r^3}\right] }{\left[1+{4 c_1 \alpha {\bar M}(r) \over r^3}\right]\left[1-{2 c_1 {\bar M}(r) \over r}\right]}, \label{e2.11d}
\end{eqnarray}
and
\begin{eqnarray}
M_{\odot}\frac{d{\bar M}(r)}{dr} = \frac{6 c_1 \alpha {\bar M}(r)^2 + c_2 r^6 {\bar \epsilon}(r)}{4 c_1 \alpha r {\bar M}(r) + r^4}, \label{mr}
\end{eqnarray}
where $c_1 \equiv \frac{G M_{\odot}}{c^2} = 1.474 \text{ km}$ and $c_2 \equiv \frac{4 \pi \epsilon_0}{M_{\odot} c^2} = 1.125 \times 10^{-5} \; \text{km}^{-3}$. The relationship between mass $M$ and radius $R$ can be straightforwardly quantified using Eq. (\ref{mr}) with a given EoS. As a result, the final two Eqs. (\ref{e2.11d}) and (\ref{mr}) can be numerically solved for a given EoS $P=P(\epsilon)$. In the next section, we will discuss the equation of state based on an interacting quark matter.

\section{Interacting Quark Matter Equation of State}\label{sec3}

The high density and relatively low temperature required to produce color superconducting quark matter may be attained in compact stars (hybrid neutron stars or strange stars).  Even though much effort to explore the EoS and other properties of matter in the interior of  such compact stars, the problem remains unsolved \cite{Lattimer:2015nhk}. 
This scenario has been corroborated by the determinations of the masses of PSR J1614-2230  \cite{Demorest:2010bx,Fonseca} and PSR J0348+0432 \cite{Antoniadis:2013pzd} have set an observational bound on the maximum mass of a NS not lower than about 2 $M_{\odot}$.
In addition to these astrophysical observations of the pulsar can be employed to constrain the composition and behaviour of the theoretical models of the EOS. There are some strange stars \cite{Bombaci:1997zz,Li,Li:1999wt} (at the moment hypothetical objects)  that can be viewed as an ultra-compact NSs (neutron stars), 
where it is possible to fit the EoS associated with these types of objects. 
Thus, the discovery of pulsars may not adjust their masses and radius to the NSs models, but set a lower limit to the maximum mass and mass-radius relation that could have led to an alternative to typical NSs. Then NSs may be converted to quark stars (QSs) 
\cite{Bombaci:2004mt,Staff:2006qf}, which consists of a deconfined mixture of
up ($u$), down ($d$) and strange ($s$) quarks (together with an appropiate number of electrons
to guarantee electrical neutrality) satisfying the Bodmer-Witten hypothesis \cite{Bodmer,Witten}. Such compact stars are referred as strange quark stars or shortly strange stars (SS).
\begin{figure}[!h]
    \centering
    \includegraphics[width = 7.5 cm]{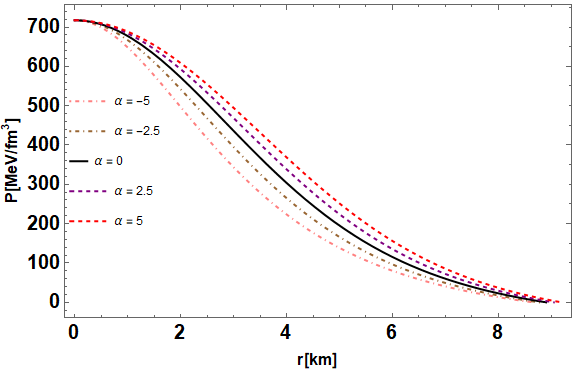}
    \includegraphics[width = 7.5 cm]{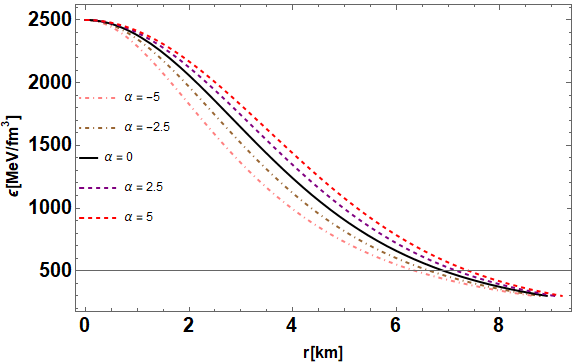}
    \caption{For the interacting EoS with different values of $\alpha = 0, \pm 2.5, \, \text{and} \, \pm 5\,$, where we set $P(r_0) = 700.00 \, \text{MeV} / \text{fm}^3, B = 70.00 \, \text{MeV} / \text{fm}^3$, we display the variation of pressure (left panel) and the energy density (right panel) with radius.}
    \label{f1}
\end{figure}
A widely accepted and easy-to-handle quark star model is the so-called thermodynamic bag model. 
The most prominent bag model is  known as the \textit{MIT bag model} \cite{Chodos:1974je},
which is the simplest and frequently used form to illustrate the interior a quark star. But, the reliable existence of the QSs, whose hypothesis cannot be conclusively ruled out
depending on the bag constant $B$, which explicitly violated the chiral symmetry of quantum chromodynamics (QCD).
Incidentally, there are many other models based on QCD corrections of second and fourth order with the aim of giving an approximate characterization of confined quarks, see \cite{Flores:2017kte}.

Here we discuss the EoS that used in modeling the strange star. The EoS  is assumed to be homogeneous and unpaired charge neutral 3-flavor interacting quark matter, which we describe using the simple thermodynamic Bag model EoS \cite{Alford:2004pf} with $\mathcal{O}$ $(m_s^4)$ corrections that account for the moderately heavy strange quark. According to Ref. \cite{Asbell:2017zxp}, an interacting quark EoS is given by
\begin{eqnarray} \label{Prad}
P&=&\dfrac{1}{3}\left(\epsilon-4B\right)-\dfrac{m_{s}^{2}}{3\pi}\sqrt{\dfrac{\epsilon-B}{a_4}}+\dfrac{m_{s}^{4}}{12\pi^{2}}\left[1-\dfrac{1}{a_4}+3\ln\left(\dfrac{8\pi}{3m_{s}^{2}}\sqrt{\dfrac{\epsilon-B}{a_4}}\right)\right],
\end{eqnarray}
where $\epsilon$ is the energy density of homogeneous quark matter (also to $\mathcal{O}(m_s^4)$ in the Bag model). Coming back to the EoS (\ref{Prad}), the mass $M$ and radius $R$ are determined by solving the TOV equations (\ref{e2.11d}) and (\ref{mr}).  To illustrate our approach in a simple setting, we consider the  boundary conditions $P(r_{0}) = P_{c}$ and $M(R) = M$,  and integrates Eq. (\ref{e2.11d}) outwards to a radius $r = R$ in which fluid pressure $P$ vanishes for $P(R)=0$. 
Corrections this one can obtain the quark star radius $R$ and mass $M = m(R)$. At this stage we set a very small numbers with initial radius $r_{0}= 10^{-5}$ and mass $m(r_{0})= 10^{-30}$
rather than zero to avoid discontinuities which appears in denominators within the equations.

\begin{figure}[!h]
    \centering
    \includegraphics[width = 7.5 cm]{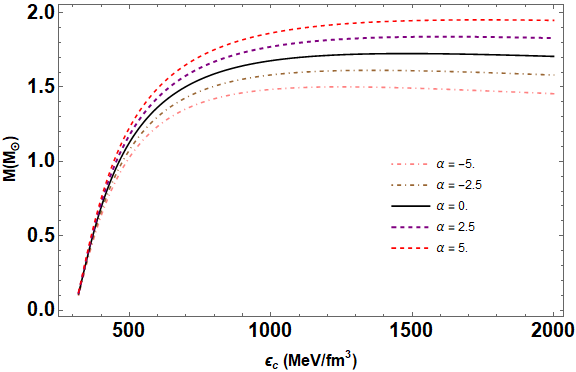}
    \includegraphics[width = 7.5 cm]{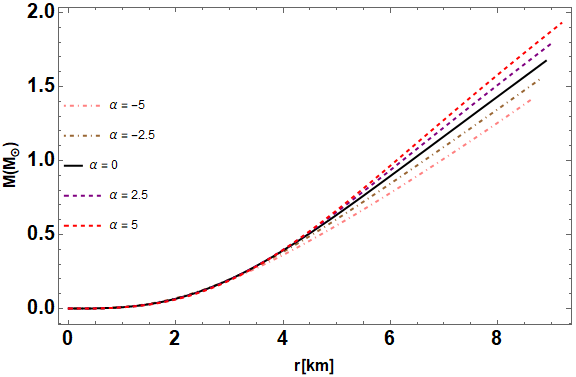}
    \caption{For the interacting EoS with different values of $\alpha = 0, \pm 2.5, \, \text{and} \, \pm 5\,$, where we set $P(r_0) = 700.00 \, \text{MeV} / \text{fm}^3, B = 70.00 \, \text{MeV} / \text{fm}^3$, we display the variation of mass versus central energy density (left panel) and the variation of mass with radius (right panel).}
    \label{f11}
\end{figure}

It is worth noting that a unit conversion $1\,{\rm fm}  = 197.3\,{\rm MeV}$ is used in order to synchronize the unit of each term given in Eq. (\ref{Prad}). Introducing this conversion, we find ${\rm MeV}^{4} = 197.3^{-3}{\rm MeV}\,{\rm fm}^{-3}$. Therefore, Eq. (\ref{Prad}) becomes
\begin{eqnarray}\label{Eos}
P &=& \frac{1}{3}(\epsilon-4B)-\frac{1}{3\pi}\sqrt{\frac{1}{197.3^{3}}}\,m^{2}_{s}\sqrt{\frac{\epsilon-B}{a_4}}\nonumber\\&+&\frac{1}{12\pi^{2}}\frac{m^{4}_{s}}{197.3^{3}}\Big[1-\frac{1}{a_4}+3\ln\Big(\frac{8\pi}{3m^{2}_{s}}\sqrt{197.3^{3}}\sqrt{\frac{\epsilon-B}{a_4}}\Big)\Big].
\end{eqnarray}
The strange quark mass $m_{s}$ will be assumed to be 100 MeV \cite{Beringer:1900zz}, and $B$ is the Bag constant whose standard accepted range is around $57\leq B \leq 92$ MeV/fm$^3$ determined by the stability 
condition with respect to iron nuclei for 2-flavour and the 3-flavour quark matter \cite{Blaschke:2018mqw}, respectively.
Finally, the parameter $a_4$ comes from the QCD corrections on the pressure of the quark free Fermi sea, which is related to the maximum mass of the star around $2M_{\odot}$ at $a_4\approx 0.7$ as suggested in \cite{Fraga:2001id}. For the study of quark matter with $\mathcal{O}(m_s^4)$ corrections, we demonstrate how the pressure and energy density do distribute by using a median value of the bag constant range such that $B\sim 70 \,{\rm MeV}/{\rm fm}^{3}$. In Fig. \ref{f1}, we plot the variation of the pressure and density with radius of the star. At that same time, we quantify the variation of mass versus  central density and the variation of mass with radius are shown in Fig. \ref{f11}. For $\alpha >0$ the mass of star for given radius increases with fixed value of $B$. In all the presented cases, one can note that there are significantly different for positive and negative values of  $\alpha$, but $\alpha =0$ case is equivalent
to standard general relativity.

\section{Numerical details and analysis of mass-radius relation}\label{sec4}

 In this section, we present the detailed results for the EoS \eqref{Eos}, and show all relevant outcomes for isotropic QSs  in the $4D$ EGB gravity. To start with, we consider a certain value of central pressure,  $P({r_{0}})=700 \,{\rm MeV}/{\rm fm}^{3}$ and the radius of the star is identified when the pressure vanishes or drops to a very small value.  Due to the long range effects of confinement of quarks, the stability of strange QSs is represented by the bag constant, $B$. We then consider the engineered TOV equations Eq. (\ref{e2.11d}) and mass function Eq. (\ref{mr}). It is important to note that the mass is measured in the solar mass unit ($M_{\odot}$), radius in ${\rm km}$, while energy density and pressure are in unit of ${\rm MeV}/{\rm fm}^{3}$. The bag constant $B$ is also in ${\rm MeV}/{\rm fm}^{3}$. In the present analysis, we treat the values of $B$ and $\alpha$ as free constant parameters. Since, the parameter $B$ can vary from $57$ to $92 \,{\rm MeV}/{\rm fm}^{3}$ \cite{Witten}. In the following, numerical values of the GB coupling $\alpha$ are given in ${\rm km}^{2}$ unit.

\begin{figure}[!h]
    \centering
    \includegraphics[width = 8 cm]{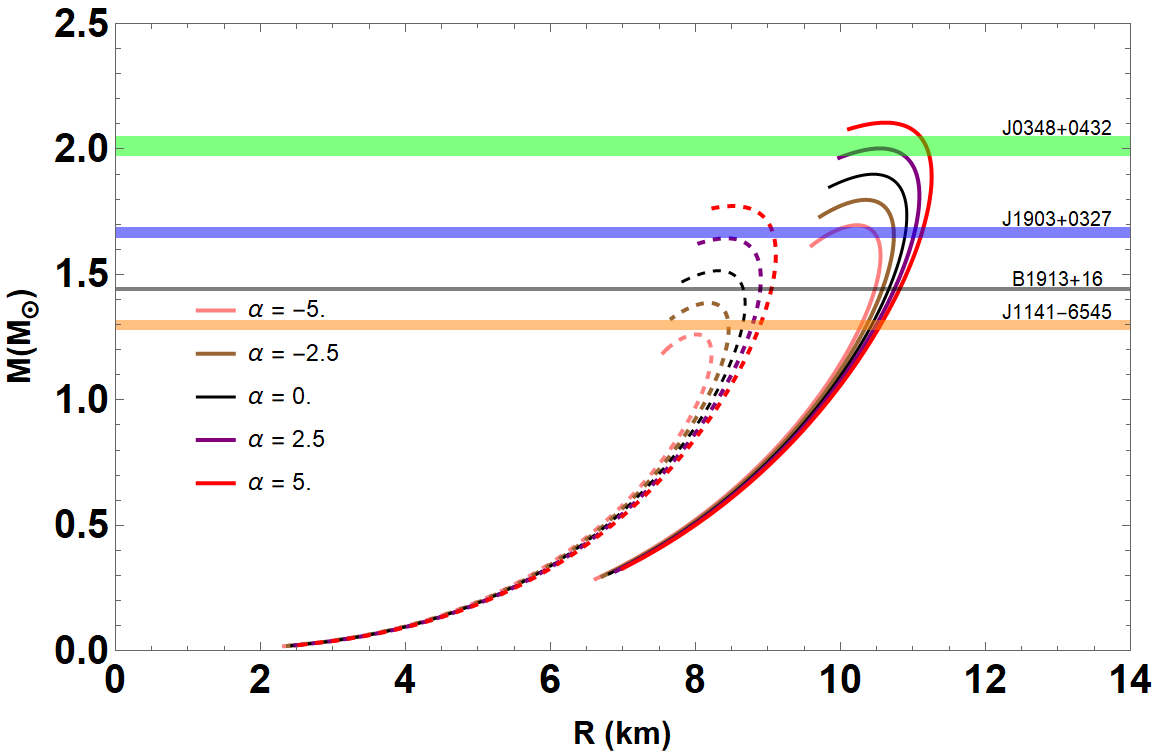}
    \caption{Figure displays the mass-radius relation
where the bag constant is set to $B = 92\,{\rm MeV}/{\rm fm}^{3}$ (dashed lines) and $B = 57\,{\rm MeV}/{\rm fm}^{3}$ (solid lines). The later represents the smallest value that the bag parameter can take. The parameters $a_{4}=0.7$ and the GB coupling $\alpha$ take several values. The horizontal bands show the observational constraints from various pulsar measurements: J0348+0432 (green) \cite{Antoniadis:2013pzd}, J1903+0327 (blue) \cite{Freire}, B1913+16 (black) \cite{Weisberg:2004hi} and J1141-6545 (orange) \cite{Bailes:2003qc}.}
    \label{f41}
\end{figure}

We depict the mass-radius curves obtained from the of the QSs as a function of the radius $R$ shown in Fig.\ref{f41} with two values of the bag constant $B$ and various values of the GB coupling $\alpha$. Moreover, in comparison our results with the data, we have used the observational constraints of the NS mass from four pulsar measurements explained in the following. The upper limit NS mass is given by Ref. \cite{Antoniadis:2013pzd} with mass $2.01 \pm 0.04 M_{\odot}$. Next, the mass from the binary pulsar J1903+0327 of $1.667\pm 0.021\, M_{\odot}$ \cite{Freire}. The NS mass is predicted $1.4408 \pm 0.008\, M_{\odot}$  form the data collection and analysis of thirty years of observations of PSR B1913+16 \cite{Weisberg:2004hi}. Lastly, the NS mass measurements of the relativistic binary pulsar PSR J1141-6545 \cite{Bailes:2003qc} is given by $1.3 \pm 0.02\,M_{\odot}$. These mass values have been utilized to compare with the mass-radius results of an anisotropic QSs with the interacting quark EoS in Ref. \cite{Becerra-Vergara:2019uzm}. In Fig.\ref{f41}, note that the mass-radius relation of the QSs in GR and $4D$ EGB cases is represented by setting $\alpha=0$ and $\alpha\neq 0$, respectively. As the results, the existence of a two solar mass compact star is found in the case when $\alpha >0 $ for the lowest values of the bag constant $B= 57 \,{\rm MeV}/{\rm fm}^{3}$, while such a star in the GR case, i.e. $\alpha=0$, can not be obtained for all possible bag constant values. On one hand, furthermore, all solutions of the TOV equation for the maximum values of the bag constant $B= 92 \,{\rm MeV}/{\rm fm}^{3}$ are located in the range of the mass constraint around $1.2\,M_{\odot} < M < 1.7\,M_{\odot}$ from pulsars J1141-6545 and J1903+0327 constrained region. On the other hand, the minimum bag constant $B= 57 \,{\rm MeV}/{\rm fm}^{3}$ gives the mass solution around $1.6\,M_{\odot} < M < 2.1\,M_{\odot}$ which are in the pulsars J1903+0327 and J0348+0432 mass constrained region. 

\begin{figure}[h!]
    \centering
    \includegraphics[width = 7.5 cm]{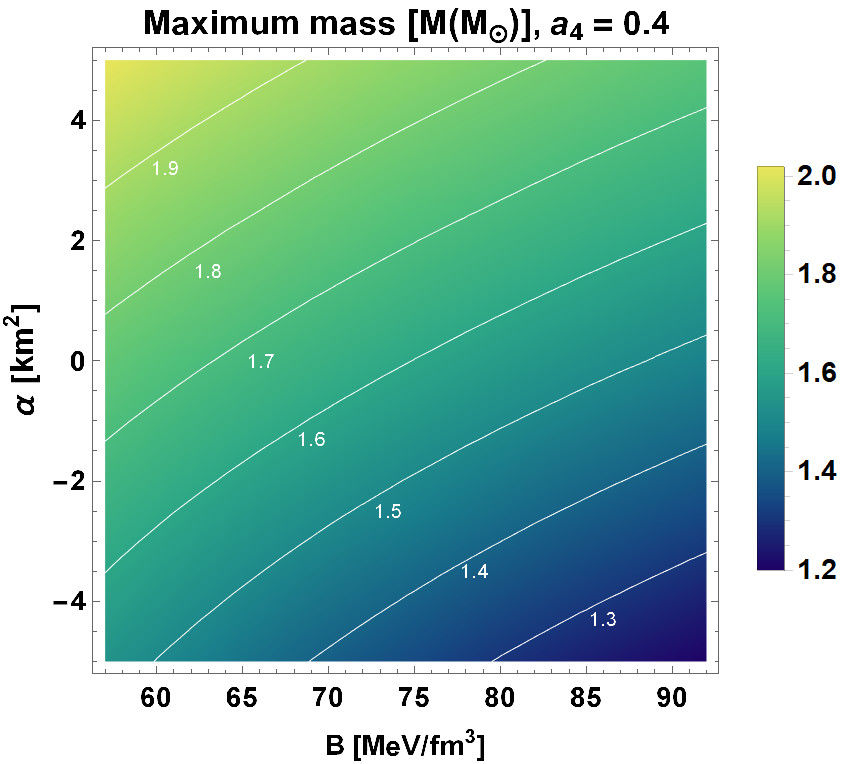}
    \includegraphics[width = 7.5 cm]{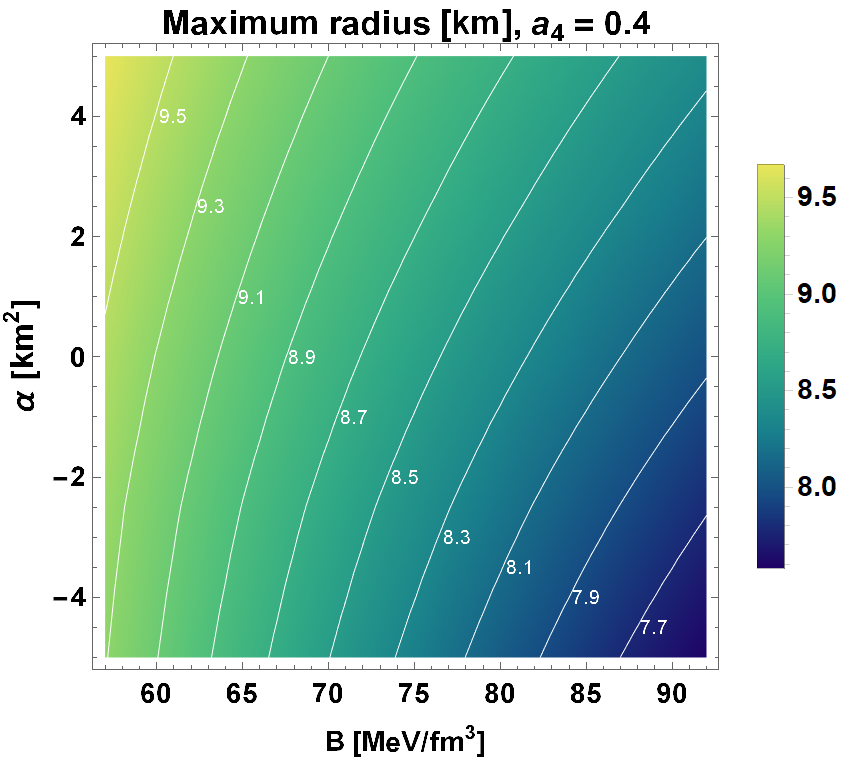}
    \caption{Figure shows the maximum masses (upper panel) and their corresponding radii (lower panel) for values of $-5\leq \alpha \leq 5$ and $57\,{\rm MeV}/{\rm fm}^{3}\leq B\leq 92\,{\rm MeV}/{\rm fm}^{3}$. We have considered a particular value of of the fourth-order-corrected parameter $a_{4}=0.4$. The white lines are equipped masses and radii lines.}
    \label{f711}
\end{figure}
\begin{figure}[h!]
    \centering
    \includegraphics[width = 7.5 cm]{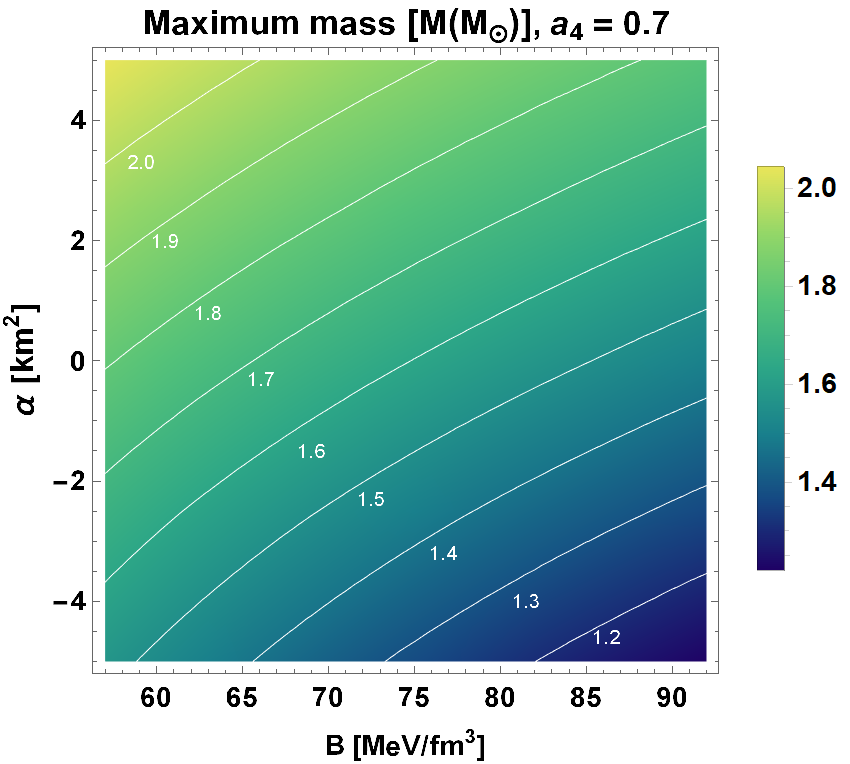}
    \includegraphics[width = 7.5 cm]{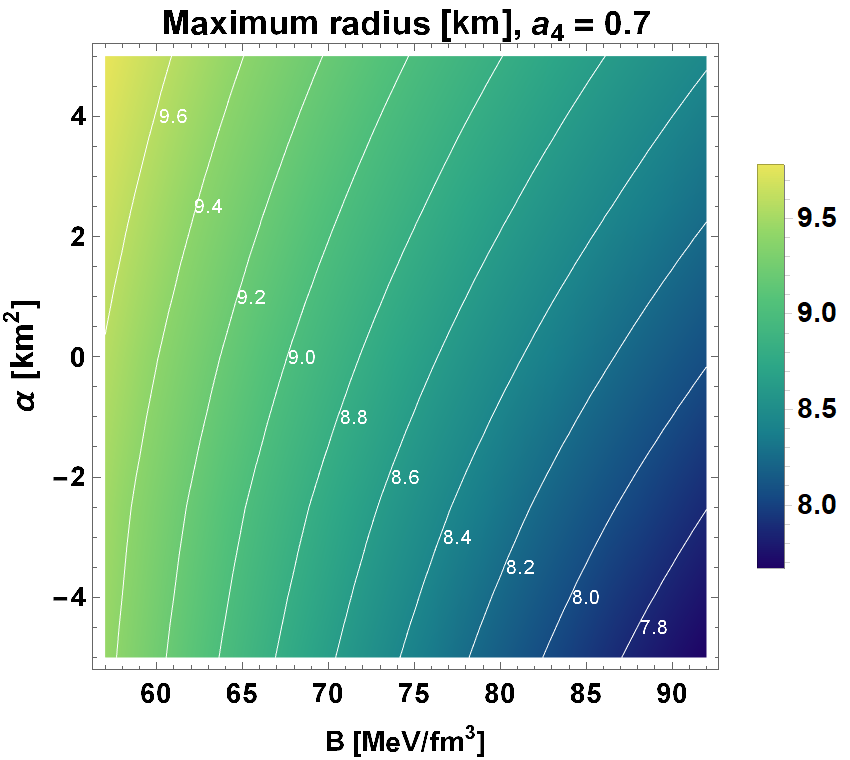}
    \caption{Maximum masses and their corresponding radii have been plotted. Same as of Fig. \ref{f711} for $a_{4}=0.7$.}
    \label{f712}
\end{figure}
\begin{figure}[h!]
    \centering
    \includegraphics[width = 7.5 cm]{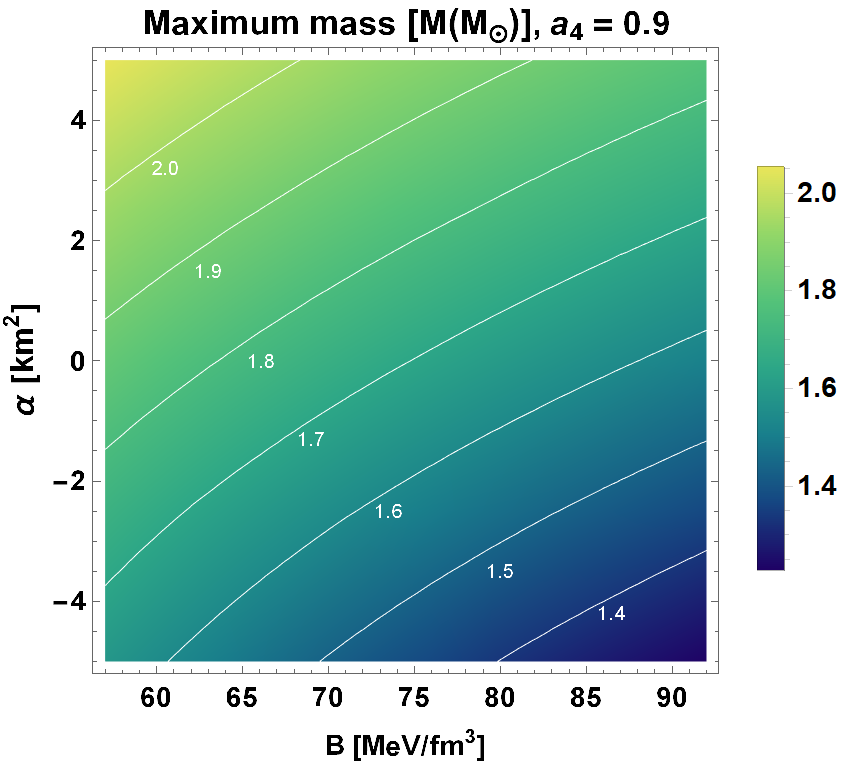}
    \includegraphics[width = 7.5 cm]{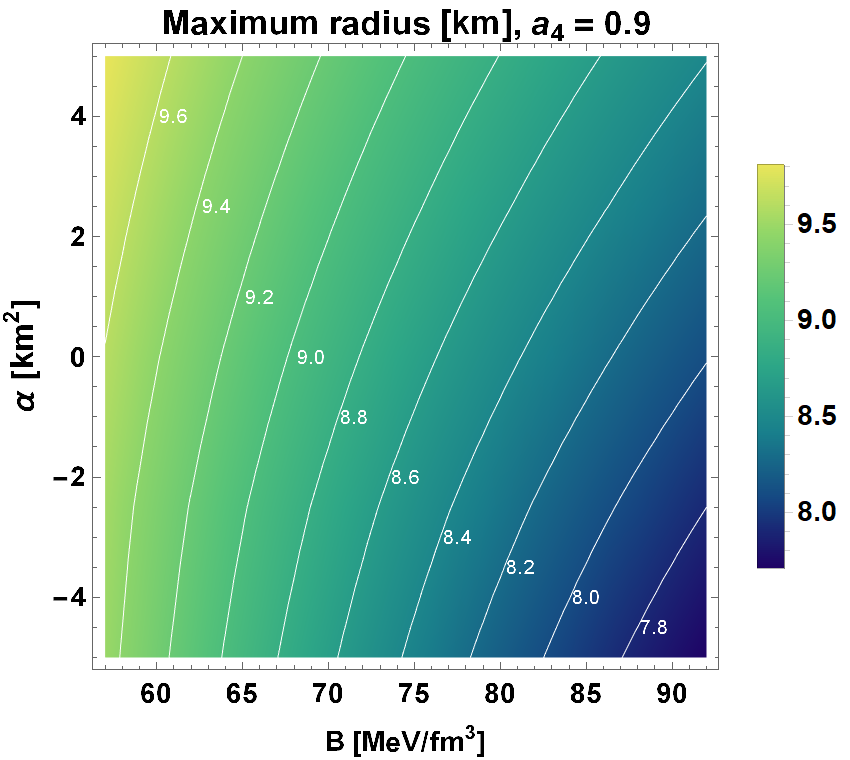}
    \caption{Maximum masses and their corresponding radii have been plotted. Same as of Fig. \ref{f711} for $a_{4}=0.9$.}
    \label{f713}
\end{figure}

We also further investigate the maximum values of the QS mass (in the solar mass unit) and its radius (km unit) from the TOV equation in the $4D$ EGB gravity with the interacting quark EoS. The numerical results can be represented in the contour plots for all possible bag constant values and the range of the GB coupling $-5\,{\rm km}^{2}\leq \alpha \leq 5\,{\rm km}^{2}$ with three values of $a_4 =0.4,\,0.7$ and $0.9$ and they are displayed in Figs. (\ref{f711}-\ref{f713}). The results show that the GB gravity coupling $\alpha$ plays an important role for enhancing or reducing the maximum mass of the QS masses as well as the radii with respect to the relative signs of the GB coupling. While, the enhancement of the bag constant reduces the masses and radii of the QSs. In addition, according to the results in the anisotropic QS case, Ref. \cite{Becerra-Vergara:2019uzm} has speculated that more interacting quarks lead to less values of the maximum masses, and vice versa. We observe that our results are also compatible with the speculation in Ref. \cite{Becerra-Vergara:2019uzm}. 

It is worth mentioning that the relevance of this theory from  astrophysical view point. Here it is important to contextualize the results in light of basic constraints on the coupling parameter. For instance in \cite{Clifton:2020xhc} authors have proposed a tightest constraints on positive values, leading to overall bounds $0 \lesssim \alpha \lesssim 10^2~ \text{km}^{2}$ based on observations of binary black holes. Our results show that, one may able to reach a maximum mass above the observed value $M_{\rm max}> 2 M_{\odot}$ for $\alpha> 3.75,\,3.25$ and $2.75\,{\rm in ~ km}^{2}$ for  $a_{4}=0.4,\,0.7$ and $0.9$, respectively. As it is seen from the Figs. (\ref{f711}-\ref{f713}) that the theoretical requirements of the model are in agreement with the current results for positive value of $\alpha$.

\section{Structural properties of strange stars }\label{sec5}

For completeness, we would also like to explore the physical properties in the interior of the  fluid sphere.

\begin{figure}[h!]
    \centering
    \includegraphics[width = 7.5 cm]{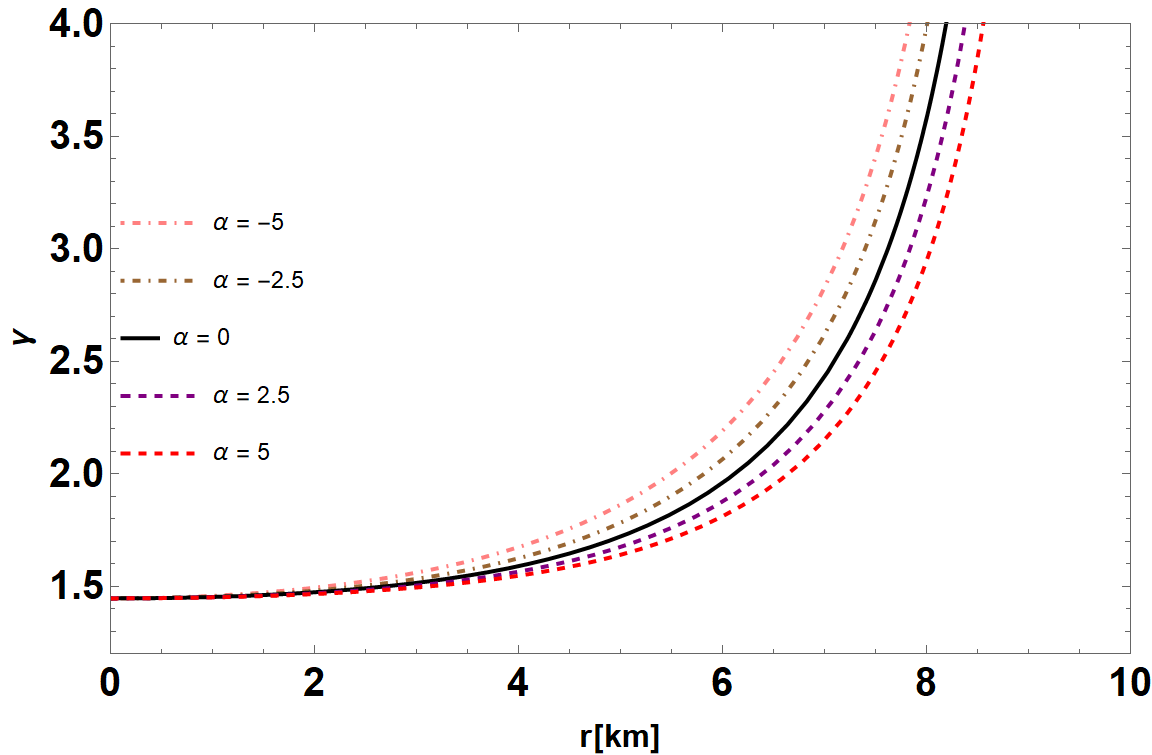}
    \includegraphics[width = 7.5 cm]{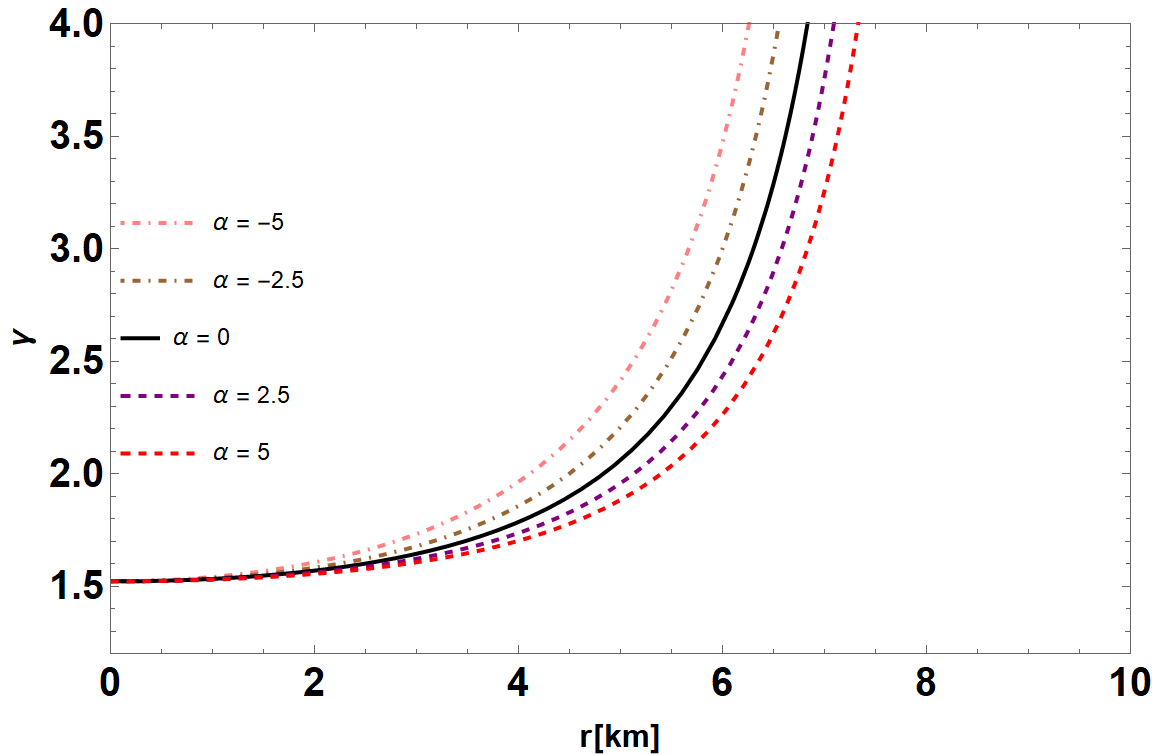}
    \caption{Plots for adiabatic index, $\gamma$, of the stars using the interacting EoS. The parameters are the same as used in Fig. \ref{f1}. }
    \label{f7}
\end{figure}

\begin{figure}[h!]
    \centering
    \includegraphics[width = 8 cm]{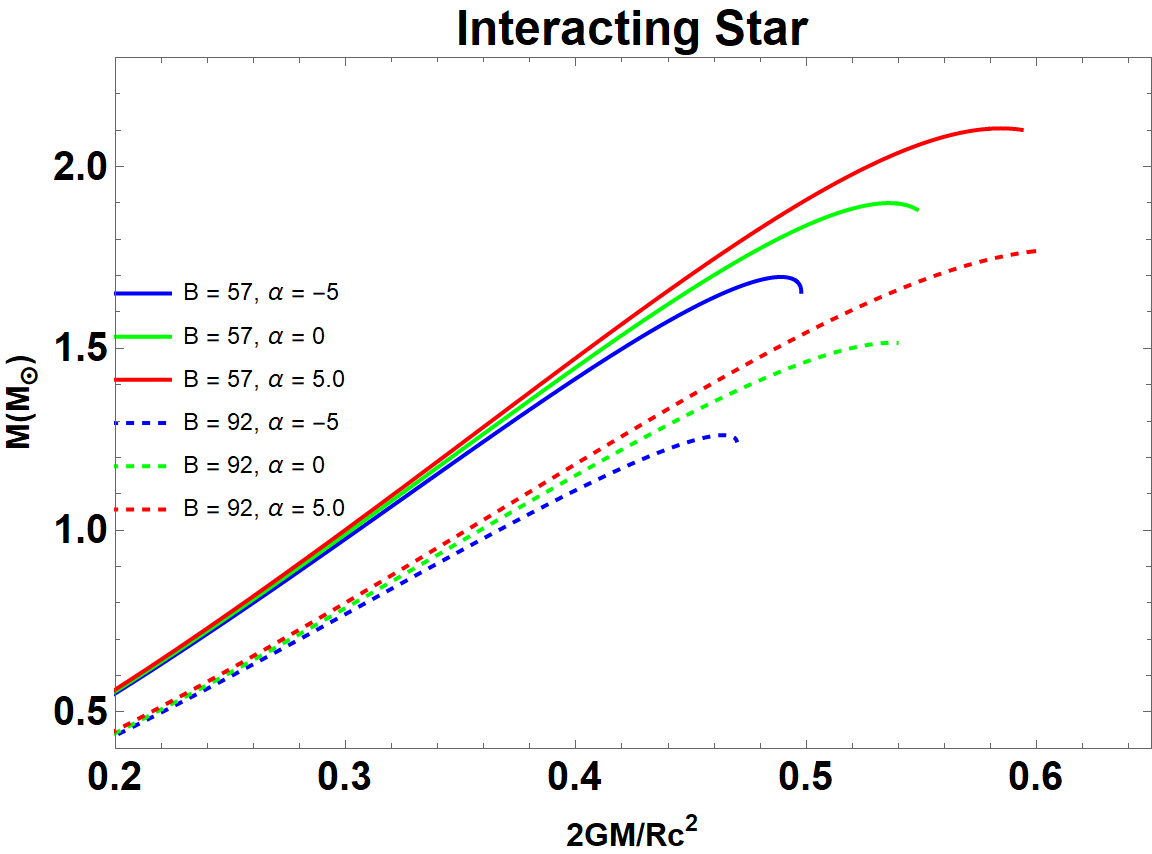}
    \caption{Compactness versus star mass for various $\alpha$ and  $B = 57, 92$ in $\, \text{MeV} / \text{fm}^3$, respectively. }
    \label{f8}
\end{figure}

\subsection{The stability criterion and the adiabatic indices }
We begin our consideration of stability in stars by examining adiabatic index ($\gamma$) based on our EoS concerning quark matter models. Since, the adiabatic index is a basic ingredient of the  instability criterion, and is related to the thermodynamical quantity. This method was introduced by Chandrasekhar \cite{Chandrasekhar} for dynamical stability based on the variational method. For an adiabatic perturbation, the adiabatic index, which appears in the stability formula, as described by the equation \cite{Chandrasekhar,Merafina}
\begin{equation}\label{adi}
\gamma \equiv \left(1+\frac{\epsilon}{P}\right)\left(\frac{dP}{d\epsilon}\right)_S,
\end{equation}
where  $dP/ d\epsilon$  is the speed of sound in units of speed of light and the subscript $S$ indicates the derivation at constant entropy. Note that the above equation is a dimensionless quantity measuring the stiffness of the EoS. 

In general, the EoS related to neutron star matter, $\gamma$ lies between  2 to 4 \cite{Haensel}. The analysis in \cite{Moustakidis:2016ndw} shows that adiabatic index on the instability conditions are also applicable 
to describe compact objects including white dwarf, neutron stars and supermassive stars. Since the value of $\gamma$ should exceed $4/3$ for relativistic polytropes depending on the ratio $\epsilon/P$ at the centre of the star \cite{Glass}. In support of  $\gamma  > 4/3$, authors in \cite{Chavanis} have found for stability of an extended cluster with $\rho_e/ \rho_0 \ll 1$ in Newtonian gravity. Finally, our results are shown in Fig. \ref{f7}, where we plot $\gamma $
as a function of radius. From Fig. \ref{f7}, the resulting $ \gamma >4/3 \sim 1.33$ shows that our model 
is stable  against the radial adiabatic infinitesimal perturbations and increasing values of $\gamma$
mean the growth of pressure for a given increase in energy density, i.e. a stiffer EoS.

\begin{table}[ht!]
\begin{center}
\begin{tabular}{|c|c|c|c|c|c|}
\hline\hline
\multicolumn{6}{l}{\hskip 3cm \mbox{Quark Stars with $B = 57$}} \\
\hline\hline
$\alpha$ & $M_{\rm max}$ &  $R$  & $\epsilon_c$ (Max) & $v_{s}/c$ & $B^{\rm max}_{\rm bind}$ \\
 {\rm km}$^{2}$ &  $(M_{\odot})$ & (km) & (MeV/fm$^3$) &   &  ($M_{\rm max}$)  \\
  \hline
  $-$5.0 & 1.697  & 10.551 & 2 $\times 10^{3}$  & 0.546 & 0.163 \\
   \quad  0 & 1.900 & 10.913 & 2 $\times 10^{3}$ & 0.546  & 0.182 \\
  \quad 5.0 & 2.105 & 11.256 & 2 $\times 10^{3}$ & 0.546  & 0.202 \\
  \hline\hline
  \multicolumn{6}{l}{\hskip 3cm \mbox{Quark Stars with $B = 92$}} \\
\hline\hline
$\alpha$ & $M_{\rm max}$ &  $R$  & $\epsilon_c$ (Max) & $v_{s}/c$ & $B^{\rm max}_{\rm bind}$ \\
 {\rm km}$^{2}$ &  $(M_{\odot})$ & (km) & (MeV/fm$^3$) &   &  ($M_{\rm max}$)  \\
  \hline
  $-$5.0 & 1.261  & 8.219 & 3.20 $\times 10^{3}$  & 0.552 & 0.155 \\
    \quad 0 & 1.515 & 8.683 & 3.20 $\times 10^{3}$ & 0.552  & 0.174 \\
  \quad 5.0 & 1.773 & 9.107 & 3.20 $\times 10^{3}$ & 0.552  & 0.210 \\
  \hline\hline
\end{tabular}
\caption{We summarize the parameters of the quark stars using various values of the 4D EGB coupling constant, $\alpha$. We show the maximum mass of the stars $M$ in a unit of the solar mass $M_{\odot}$ with their radius $R$ in ${\rm km}$ and the central energy density $\epsilon_c$.}\label{ta11}
\end{center}
\end{table}

\subsection{Compactness and Binding energy}  

Our next step is to calculated the emission produced by the photons from the star surface through the gravitational redshift \cite{Haensel}
\begin{eqnarray}
Z_{\rm surf}=\left(1-r_g/R\right)^{-1/2}-1,
\end{eqnarray}
where $r_g= 2 GM/c^2$, and $R$ is the radius of the star.  From this point of view,  compactness $2MG/Rc^{2}$ leads to the redshift value for the given EoS \eqref{Prad}. For clarity, we display the \textit{compactness parameter} $r_g /R$, where $r_g =2 G M/c^2$ in Fig. \ref{f8}  for particular values of the bag parameter $B = 57 ~\text{and}$  $ 92 \, \text{MeV} / \text{fm}^3$ with different GB coupling constant $\alpha= 0,\pm\,5\,{\rm km}^{2}$.

As pointed in Ref. \cite{Haensel}, there exists a
universal relation between the total binding energy and the stellar mass of the neutron star. The binding energy ($B_{\rm bind}$) of a stable neutron star correlates with its gravitational mass. A more precise formula of the binding energy, containing the compactness parameter $\beta =
r_{g}/R$, was proposed by Lattimer \& Prakash \cite{Lattimer:2000nx}. It is formulated via the following relation:
\begin{eqnarray}
B_{\rm bind}\simeq 1.6\times 10^{53}\Big(\frac{M}{M_{\odot}}\Big)\Big(\frac{\beta}{0.3}\Big)\frac{1}{1-0.25\beta}\,{\rm erg}.
\end{eqnarray}
In terms of the radius dependence, it is given by
\begin{eqnarray}
\frac{B_{\rm bind}}{M}\simeq \frac{0.298\,\beta}{1-0.25\beta}.
\end{eqnarray}
An approximated value of $B_{\rm bind}=B^{\rm max}_{\rm bind}$ for $M=M_{\rm max}$ is shown in the last column of Table \ref{ta11} for its corresponding radius, $r=R$.

\section{Conclusions}
\label{sec6}
In this article, we have theoretically constructed the ultra-dense compact objects called `neutron stars'. The recent discovery of pulsars by radio telescopes and X-ray satellites has imposed restrictions on the EoS that need to describe matter inside compact objects. Here, we represent the so-called quark stars by considering quark matter EoS in the context of recently proposed $4D$ Einstein-Gauss-Bonnet gravity. As mentioned in the introduction that
$4D$ EGB theory does not have the standard  field equations, and thus we started from regularized $4D$ EGB gravity. In next we show that the trace of the field equations (\ref{GB3}) is exactly  same form as the trace of the field  equations for novel $4D$ EGB theory in a static and spherically symmetric spacetime, while the regularized
version could bypass the above issues.

Astronomical observations in favour of the possible existence of compact stars are partially or totally made up of quark matter. But the existence of quark stars is still controversial and its EoS is also uncertain. Here, we have solved the TOV equation in the $4D$ EGB gravity with the interacting quark EoS. To be more specific, we have studied millisecond pulsars modelled as quark stars with interacting quark EoS. As the results, several solutions of the mass-radius relation are compatible with four pulsar constraints of the NS mass with upper and lower limits of the bag constant values. We obtained the two solar mass NSs in the $4D$ EGB gravity with the lowest value of the bag constant while it is not possible for the standard GR theory. More importantly, increase and decrease of the QS masses are controlled by the plus and minus signs of the GB coupling, $\alpha$, respectively where as the enhancement of the bag constant reduces the NS masses. In addition, we further investigate the maximum values of the masses and radii of the QSs by varying the GB gravity coupling and the bag constant with the three values of the $a_4$. We found that more interacting quarks reduce the maximum mass of the QSs, and vice versa. On one hand, furthermore, the stability of the QSs can be achieved by the given values of the parameters in the theory. On the other hand, the binding energies of the QSs in $4D$ EGB gravity are calculated and they provided reasonable values for the observation data.

In the conclusion, the novel $4D$ EGB gravity provided a good result in the analysis of the mass of the QSs. In particular for the two solar mass of the observed NSs, it has been well known that it is difficult to obtain the two solar mass of the NSs in the standard GR gravity with several models of the EoS. However, the appearance of the $4D$ EGB gravity come to rescue for this problem. The mass of the NSs or QSs can be increased or decreased depending on the magnitudes of the higher order gravity coupling as shown in this work. To gain more and deeper understanding of the compact objects in the framework of the $4D$ EGB gravity, an extended analysis is worth for further study related to astrophysical observables, for instances, gravitational waves signal from binary system, pulsar timing array, accretion disk analysis of the NSs and \textit{etc}. We leave them for further investigation.

\section*{Acknowledgments}
The authors are grateful to the referee for careful reading of the paper and valuable suggestions and comments.  T. Tangphati would like to thank the financial support from the Science Achievement Scholarship of Thailand (SAST). P. Channuie acknowledged the Mid-Cereer Research Grant 2020 from National Research Council of Thailand under a contract No. NFS6400117.

\end{document}